\documentclass{acm_proc_article-sp}

\usepackage{graphicx}
 \newtheorem{definition}{Definition}

 \makeatletter
\newif\if@restonecol
\makeatother

\usepackage{algorithm2e}

\begin{document}

\title{Distributed Area Coverage by Connected Set Cover Partitioning in Wireless Sensor Networks}
\numberofauthors{2}
\author{
\alignauthor
Dibakar Saha \\
      \affaddr{Advanced Computing and Microelectronics Unit}\\
       \affaddr{Indian Statistical Institute, Kolkata, India}\\
       \email{dibakar.saha10@gmail.com}
\alignauthor
Nabanita Das \\
       \affaddr{Advanced Computing and Microelectronics Unit}\\
     \affaddr{Indian Statistical Institute, Kolkata, India}\\
       \email{ndas@isical.ac.in}
       }
     
\maketitle

\begin{abstract}
Assuming a random uniform distribution of $n$ sensor nodes over a virtual grid, this paper addresses the problem of finding the maximum number of
connected set covers each ensuring $100\%$ coverage of the query region. The connected sets remain active one after another in a round robin fashion such that if there are $\cal P$ such set covers, it can enhance the network lifetime
$\cal P$-fold. From graph-theoretic point of view, a centralized {\it O}$(n^3)$ heuristic is proposed here to maximize $\cal P$. Next, for large self-organized sensor networks, a distributed algorithm is developed. The proposed algorithm is to be executed just once, during the initialization of the network. In case of failure, a distributed recovery algorithm is executed to rearrange the partitions. Simulation studies show that the performance of the proposed distributed algorithm is comparable with that of the centralized algorithm in terms of number of partitions. Also, comparison with earlier works shows significant improvement in terms of number of partitions, message complexity and network lifetime. 

\end{abstract}
\section{Introduction}
\label{sec_1}
Wireless sensor network (WSN) has recently gained popularity in many applications like health care, defense, security, environment monitoring etc. Each sensor node of a WSN is equipped with processing elements, sensors, transceivers and limited memory. Also, the energy source of a sensor node is limited due to its small dimension. In addition, once deployed, a sensor network usually works for several weeks or months without any maintenance. Hence, if a node relinquishes its energy the network fails to work. To make the network robust against such failures, WSN's are usually over deployed. In such an over deployed WSN, the lifetime can be enhanced, if instead of keeping all the sensors awake always, only a subset of sensor nodes is kept active at a particular instance that is sufficient to cover the region. Also, the information gathered by the sensor nodes in any subset should be finally transferred to the sink node via multi-hop path. Therefore, the active set of sensors must be connected. The remaining 
sensor nodes which are not in any subset can be used as backup nodes to protect the network against faults. It is evident that given a set $\cal N $ of sensor nodes distributed over a query region $\cal Q $, if we can partition $\cal N $ into $\cal P $ subsets such that the nodes in each partition are connected and are sufficient to cover $\cal Q $, the subsets can be made active one after another in a round robin fashion to result a $\cal P$-fold increase in network lifetime. Hence, our aim is to maximize the number of such subsets satisfying coverage and connectivity requirements so that  we can achieve maximum possible network lifetime.
A lot of research activities have been reported so far on coverage and connectivity problems in WSN, formulated in various ways to combat their inherent hardness. In some works, authors considered the coverage problem only without taking into account the connectivity issue \cite{Zabrams, Huang,Meguerdichian, SSlij,Demin}. Coverage and connectivity issues together have been addressed in some recent literature. Given a random node distribution over a query region the problem of finding a connected set cover of minimum size is {\it NP-hard} \cite{Wang}. Either centralized heuristics or approximation algorithms and their distributed versions are proposed for finding a single connected set cover \cite{Himangsu}, \cite{Zhou}, where the rest of the nodes remains unused, or some dynamic scheduling algorithms are developed where nodes periodically sense their neighborhood and decide its role to achieve guaranteed degrees of coverage and connectivity \cite{Gallais,Lin,Chong,Tian,Wang,yan}.\\
$~~~~~$In most of these works, either the problem of finding a single connected $K$-cover set has been addressed without considering the connectivity requirement, or even in the cases, where both coverage and connectivity 
issues are considered, either a single connected cover is generated keeping the remaining nodes unutilized, or dynamic scheduling techniques are presented that require the nodes to waste energy in message communication to probe its neighborhood in every cycle to take the necessary action. This will incur large message overhead in terms of energy in the radio circuitry of the sensor nodes. To avoid this overhead, the authors of \cite{Pervin} propose a localized algorithm to be executed just during initialization to find the maximum number of possible connected set covers. However, in \cite{Gallais,Pervin,Tian}, the monitoring area has been assumed to be a dense grid \cite{Wei}, \cite{RSS} composed of unit cells. The knowledge of exact location of each node is needed here. Area 
coverage or overlapped areas of sensor nodes are calculated by counting the grid points for each neighbor. Sensor nodes need to memorize all the grid points to compute covered and uncovered cells. Hence even the localized algorithms are computation intensive and need lots of local memory.\\
In this paper, assuming the query region to be a virtual grid composed of a limited number of blocks, an {\it O}$(n^3)$ centralized heuristic and a distributed algorithm are developed to maximize the number of partitions of sensor nodes such that each partition is connected and offers $100\%$ coverage. The proposed distributed algorithm is executed just once during the initialization of the network without the information of the exact locations of the nodes. In case of faults, a distributed fault-recovery algorithm is developed for rearrangement of the faulty partition. Simulation studies show that the proposed distributed algorithm performs better in terms of number of partitions, percentage of active nodes and message overhead compared to that of \cite{Gallais} and \cite{Pervin}. Moreover, it improves the lifetime of the network significantly when compared to the dynamic algorithm of \cite{Gallais}.\\
Rest of the paper is organized in the following way: section \ref{sec_2} formulates the problem with necessary preliminaries, section \ref{sec_3} presents the algorithms, section \ref{sec_4} discusses the simulation details and the results, and finally section \ref{sec_5} concludes the paper.
 \section{{ Problem Overview}}
\label{sec_2}
 In this paper, it is assumed that $n$ sensor nodes are deployed randomly over a $2$D-plane termed as the query region $\cal Q$. The sensor nodes are homogeneous having the same sensing range ($S$) and transmission range ($T$). It is to be mentioned that in case, $T \geq 2S$, for any convex region, coverage ensures connectivity \cite{Wang}. In our model, we assume the more general case where $T$ and $S$ may have any value irrespective of each other.
 \subsection{Preliminaries and Proposed Model}
 This subsection introduces some notations and the system model that will be used throughout this paper.
 \vspace{-0.5cm}
 
        \begin{definition} 
     Given a set of sensors {\cal N}={\{$N_1$, $N_2$,\ldots, $N_n$\}} distributed over $\cal Q$, the {\it communication graph} ${G}({\cal N},E)$ for the sensor network is the undirected graph where an edge $(u, v) \in E$ exists if and only if the Euclidean distance between $u$ and $v$ denoted by $d(u, v) \leq T$. The $communication$ $graph$ $induced$ by a set of sensors $M \subseteq \cal N$ is the induced subgraph of ${G}({\cal N},E)$ involving only the vertices in $M$ and the corresponding edges.
    \end{definition}
    \vspace{-0.7cm}
        \begin{definition}
     Consider a sensor network consisting of a set $\cal N$ of $n$ sensors and a query region $\cal Q$. A set of sensors $M \subseteq \cal N$ is said to be a {\it connected K-cover} for the query region if, each point $p\in \cal Q$ is covered by at least $K$ sensors from $M$ and the communication graph induced by $M$ is connected.
    \end{definition}
    \vspace{-0.5cm}
    
\begin{figure}[ht]
\centering
 \includegraphics[scale=0.25]{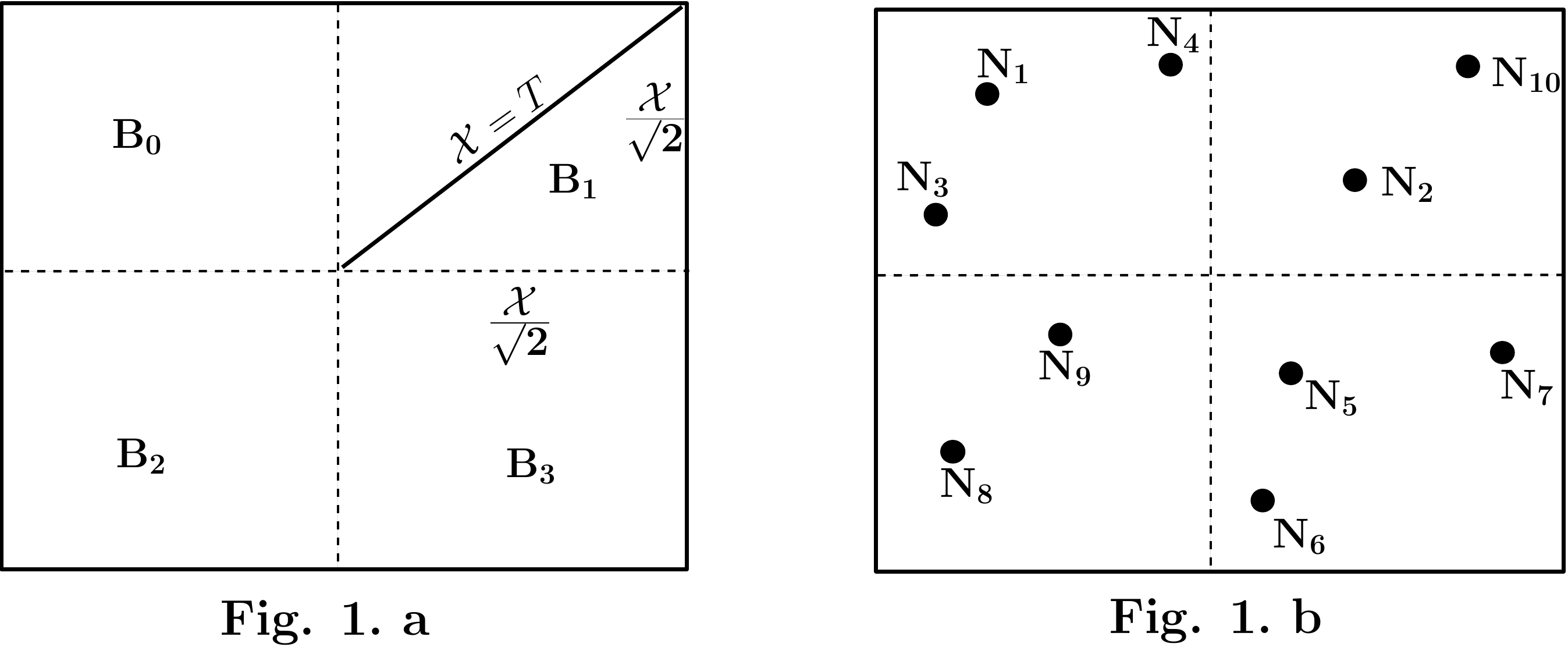}
\caption{\scriptsize{a)${\cal Q}$ divided into blocks; b)node deployment in ${\cal Q}$}}
 \label{fig1}
 \vspace{-0.1cm}
\end{figure}

\begin{figure}[ht]
\hspace{0.6cm}
\begin{minipage}[b]{0.38\linewidth}
\centering
 \includegraphics[scale=0.22]{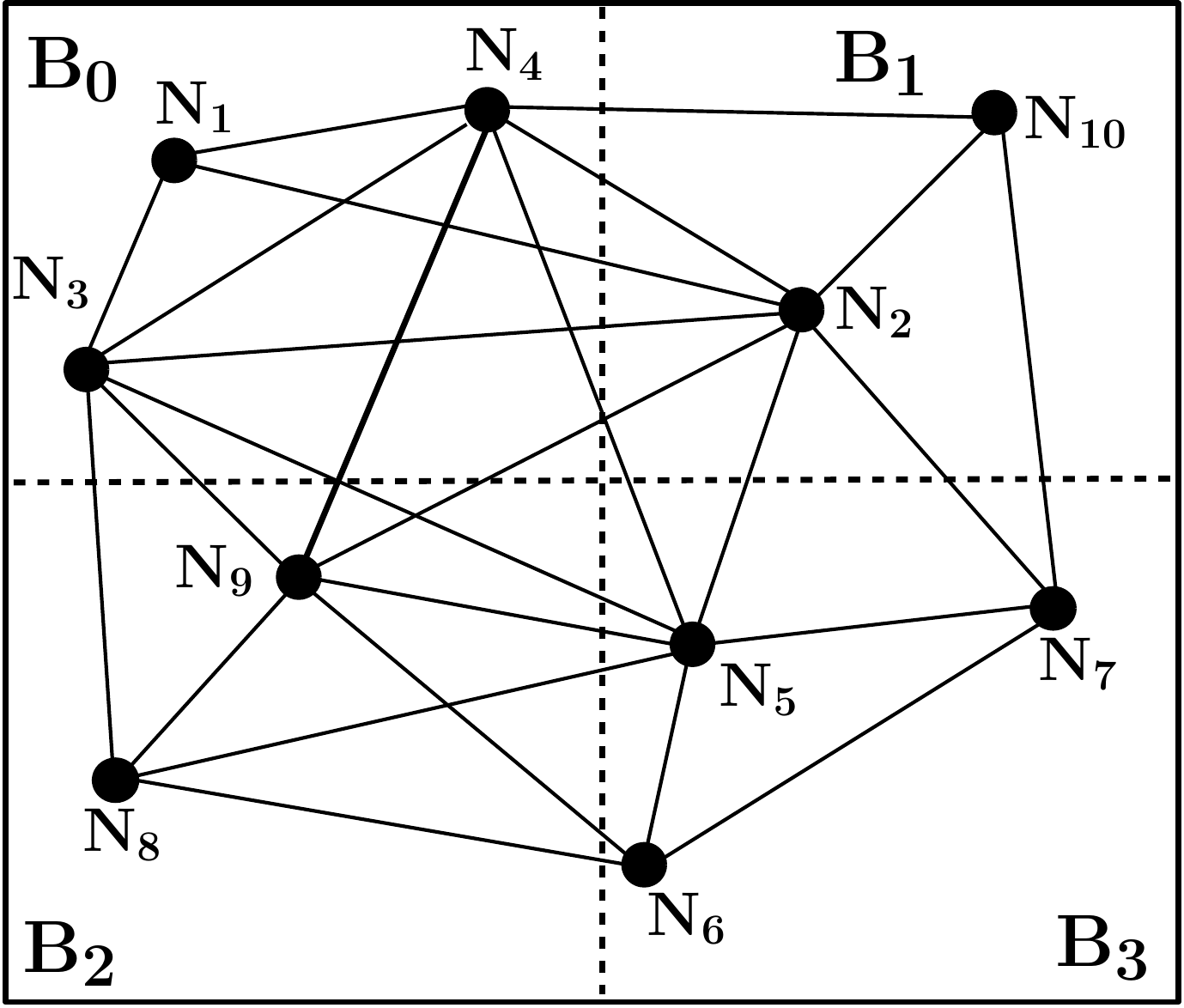}
\caption{\scriptsize{Communication graph ${{\it G}}({\cal N}, E)$ }}
 \label{fig2}
\end{minipage}
\hspace{0.5cm}
\begin{minipage}[b]{0.38\linewidth}
\centering
 \includegraphics[scale=0.23]{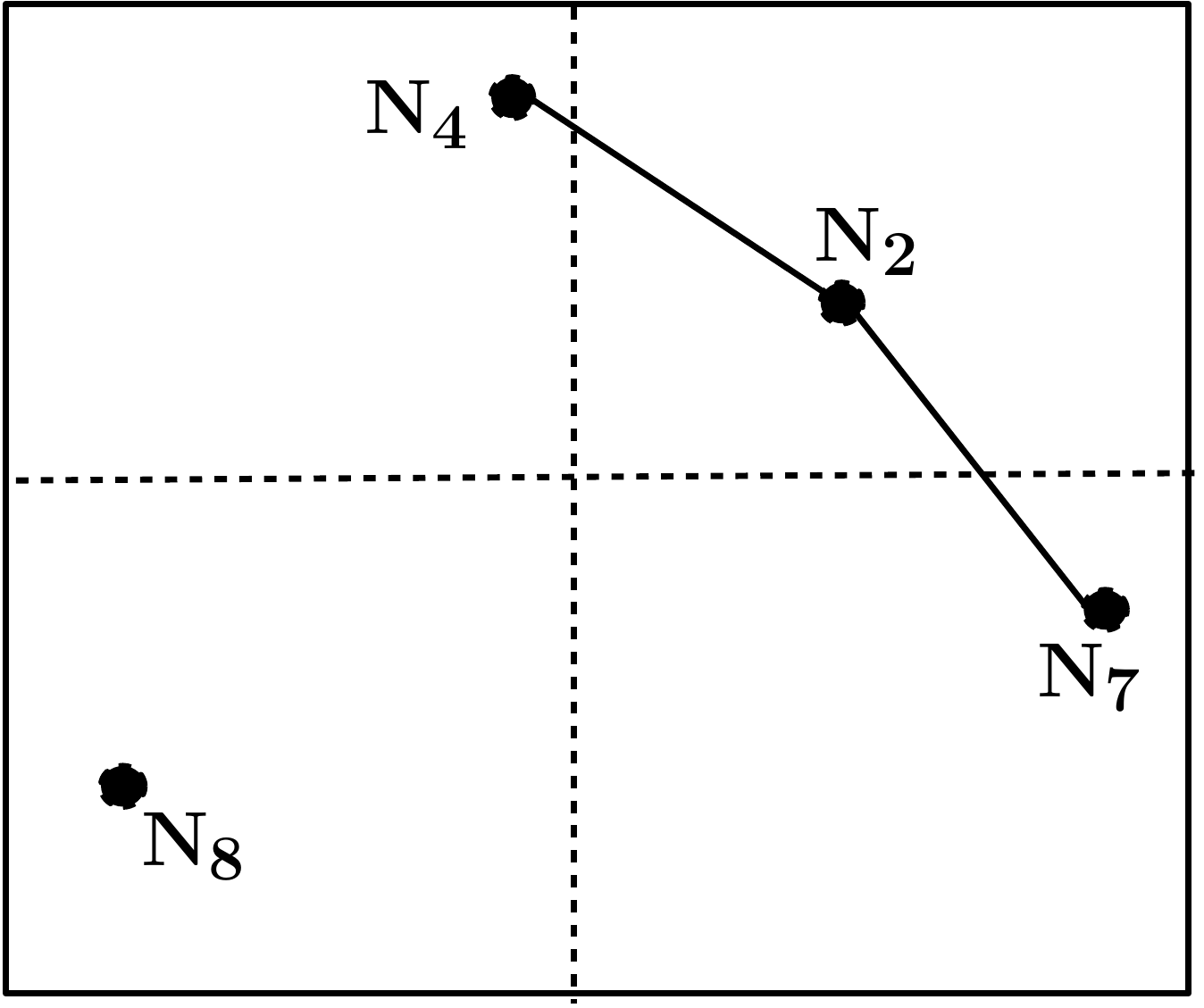}
\caption{\scriptsize{An induced subgraph of ${{\it G}}({\cal N}, E)$ }}
\label{inducedgraph}
\end{minipage}
\end{figure}

Let a set of $n$ sensor nodes {\cal N}={\{$N_1$, $N_2$,\ldots, $N_n$\}} be deployed over a 2-D query region $\cal Q$. It is assumed that $\cal Q$ is divided into a grid of square blocks $ { \cal B } = \{ B_1, B_2, ..., B_m \}$ such that the area of each block is $ \frac{{\cal X}^2}{2}$, where ${\cal X}= min( S, T)$. Here, each node knows its location only in terms of the blocks, i.e. just which block it belongs to. It is evident that all sensor nodes within the same block are connected to each other and each sensor within a block always covers the block fully. Therefore, activating at least one sensor node from each block $B_i \in \{ \cal B \}$, for $ 0 \leq i \le m$, ensures $100\%$ coverage of $\cal Q$. However, it is not a sufficient condition for connectedness of the selected subset of nodes which is in fact dictated by the underlying communication graph. 
 \vspace{-0.5cm}
\begin{paragraph}
 {\bf Example} Given a random node distribution over the 2-D plane ${\cal Q }$ divided into four blocks shown in Figure \ref{fig1}, the corresponding communication graph is shown in Figure \ref{fig2}. It is to be noted that if we select a subset of nodes, say $ \{N_4, N_2, N_7, N_8\}$ the subset is sufficient to cover ${\cal Q}$ but the induced subgraph as shown in Figure \ref{inducedgraph} is a disconnected one.\\
\end{paragraph}
\vspace{-0.5cm}
\vspace{-0.7cm}
 \begin{definition}Given a sensor network consisting of a set of sensors ${\cal N}=\{ N_1, N_2, \ldots, N_n\}$ distributed over a 2-D plane ${\cal Q }$ divided into a virtual grid of blocks, the $cell\ graph$ $G_{\cal B}({\cal B}, E_{\cal B})$ is defined as a graph where a vertex ${B_i}$ represents the collection of nodes within the block $B_i$ and two blocks are adjacent if there are two nodes, one in each block, which are adjacent in the communication graph ${{\it G}}({\cal N}, E)$ \cite{Arvind}.
 \end{definition}
  \vspace{-0.45cm}
 In this paper, we introduce the concept of a {\it weighted cell graph}.
  \vspace{-0.4cm}
 \begin{definition}
  Weighted cell graph is a cell graph where each edge $(B_i, B_j)$ has a weight $W_{ij}$ which is the number of disjoint edges $(N_i, N_j)$, existing in ${{\it G}}({\cal N},E)$, such that $N_i \in B_i$ and $N_j \in B_j$. 
 \end{definition}
\vspace{-0.5cm}
Given the random node distribution over ${\cal Q }$ shown in Figure \ref{fig1}, the corresponding weighted cell graph is shown in Figure \ref{fig4}.
\subsection{Problem Formulation} 
The goal of this paper is that given a random uniform distribution of a set of $n$ nodes over a 2-D plane, to maximize the number of partitions where each partition is a {\it connected 1-cover}. The problem is defined as the {\it Connected Set Cover Partitioning Problem} in \cite{Pervin}. Activating the {\it connected 1-covers} in a round robin fashion we can enhance the network lifetime significantly. This problem seems to be more complex than the {\it Connected Set Cover} problem i.e., to find a single {\it connected 1-cover} of smallest size which is reported to be NP-hard \cite{SSlij}. Our objective is to develop a light weight distributed protocol to be executed just once during the initialization of the network to reduce the communication overhead required for the dynamic protocols where nodes decide their roles depending on the periodic observation of their neighborhood.

Under the proposed model, the {\it Connected Set Cover Partitioning Problem} reduces to the problem of extracting maximum number of spanning trees of the {\it weighted cell graph}, as has been described in the following section. Based on this concept, a centralized algorithm and its distributed version are proposed in the following section. 
\section{Proposed Algorithms for Connected Set Cover Partitioning}
\label{sec_3}
\vspace{0.3cm}
Firstly we propose a centralized algorithm for connected set-cover partitioning problem assuming that the global network information is available to a central node. The details of the algorithm are presented below.
\vspace{-0.3cm}
  \subsection{\bf Centralized Algorithm: CCSP}
Given the set of nodes ${\cal N}$ distributed over the query region ${\cal Q}$ divided into a grid of blocks ${\cal B}$, we assume that the communication graph $G(N,E)$ is known to the network. From that to generate the maximum number of {\it connected 1-covers} the following steps are executed. 

\begin{figure}[ht]
		\begin{minipage}[b]{0.26\linewidth}
		 \centering
		   \includegraphics[scale=0.17]{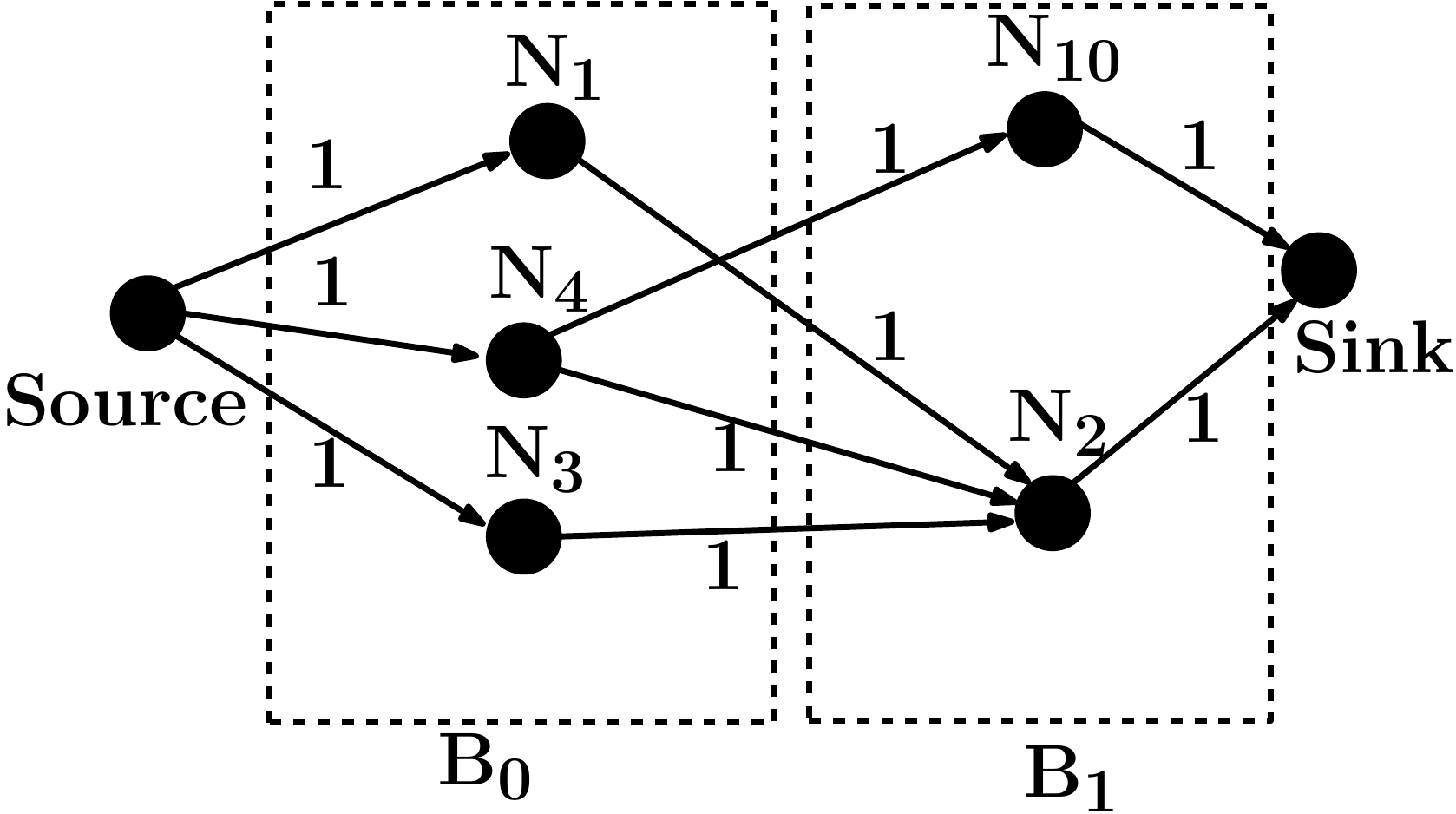}\\
		    \caption{\scriptsize{Maximum Flow Graph for blocks $B_0\&B_1$}}
		  \label{fordfulkerson}
		   \end{minipage}
	      \hspace{0.55cm}
	      \begin{minipage}[b]{0.25\linewidth}
		  \centering
		    \includegraphics[scale=0.19]{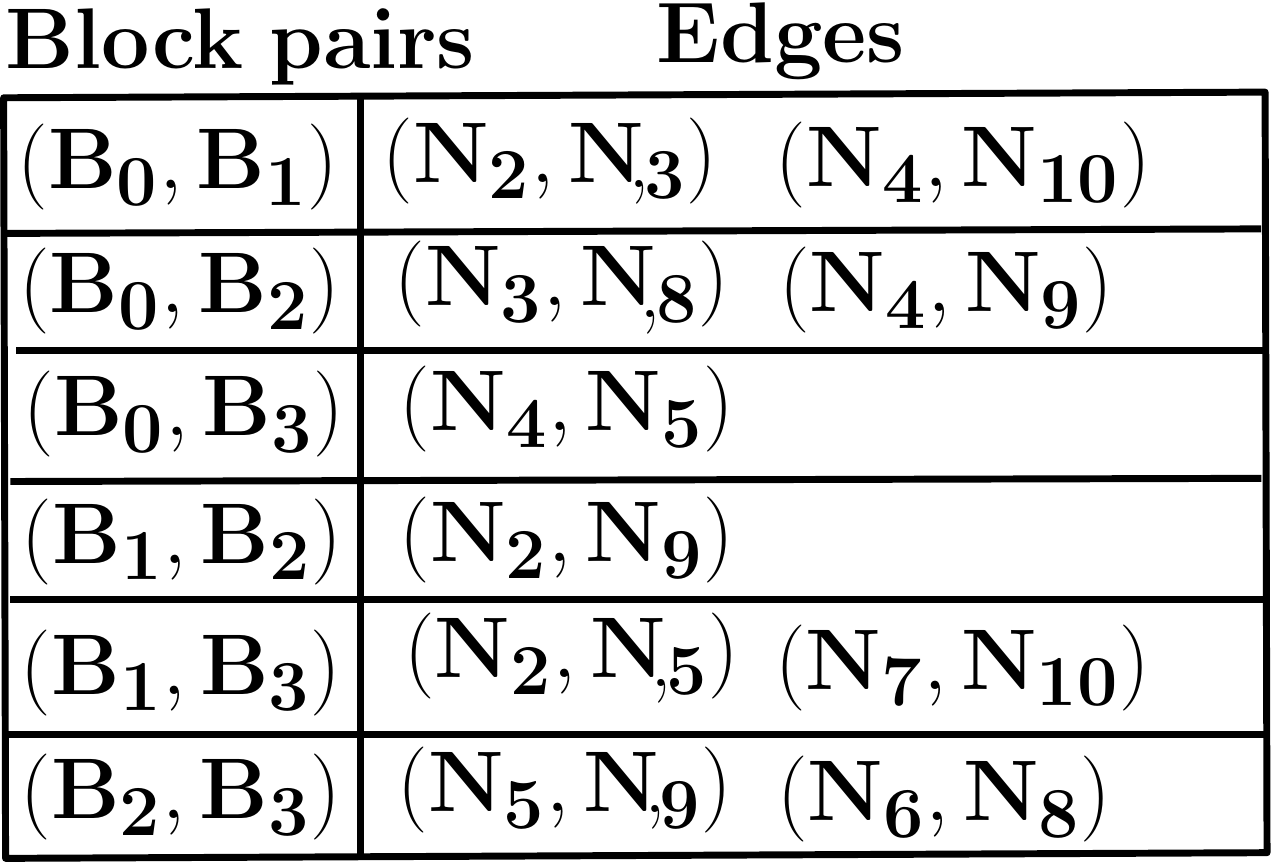}
		  \caption{\scriptsize{Disjoint edge list for all pairs of blocks}}
		  \label{fig3}
		\end{minipage}
		\hspace{0.6cm}
	      \begin{minipage}[b]{0.21\linewidth}
		  \centering
		     \includegraphics[scale=0.22]{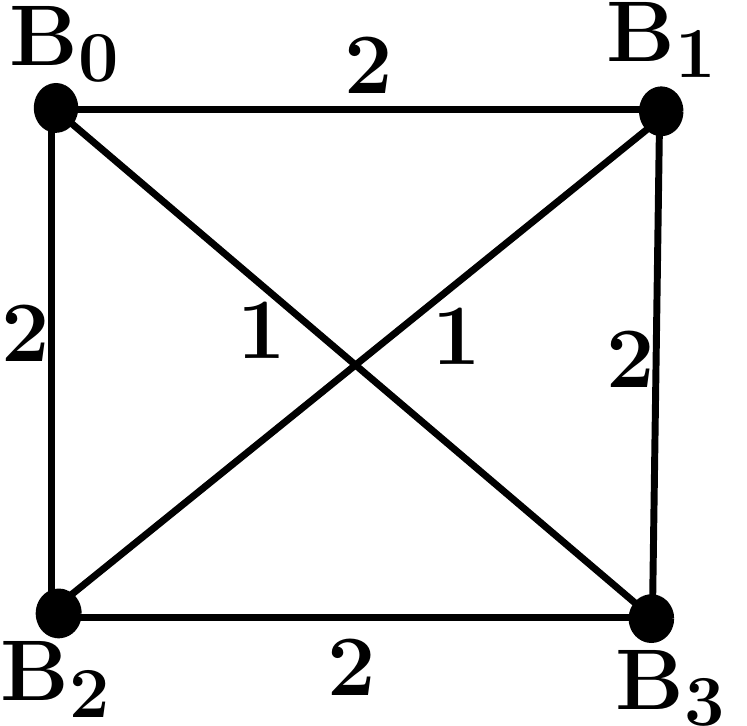}
		  \caption{\scriptsize{Weighted Cell Graph $G_{\cal B}({\cal B},E_{\cal B})$ }}
		  \label{fig4}
		\end{minipage}
		\end{figure}

\begin{figure}[ht!]
		\begin{minipage}[b]{0.5\linewidth}
		  \centering
		 \includegraphics[scale=0.23]{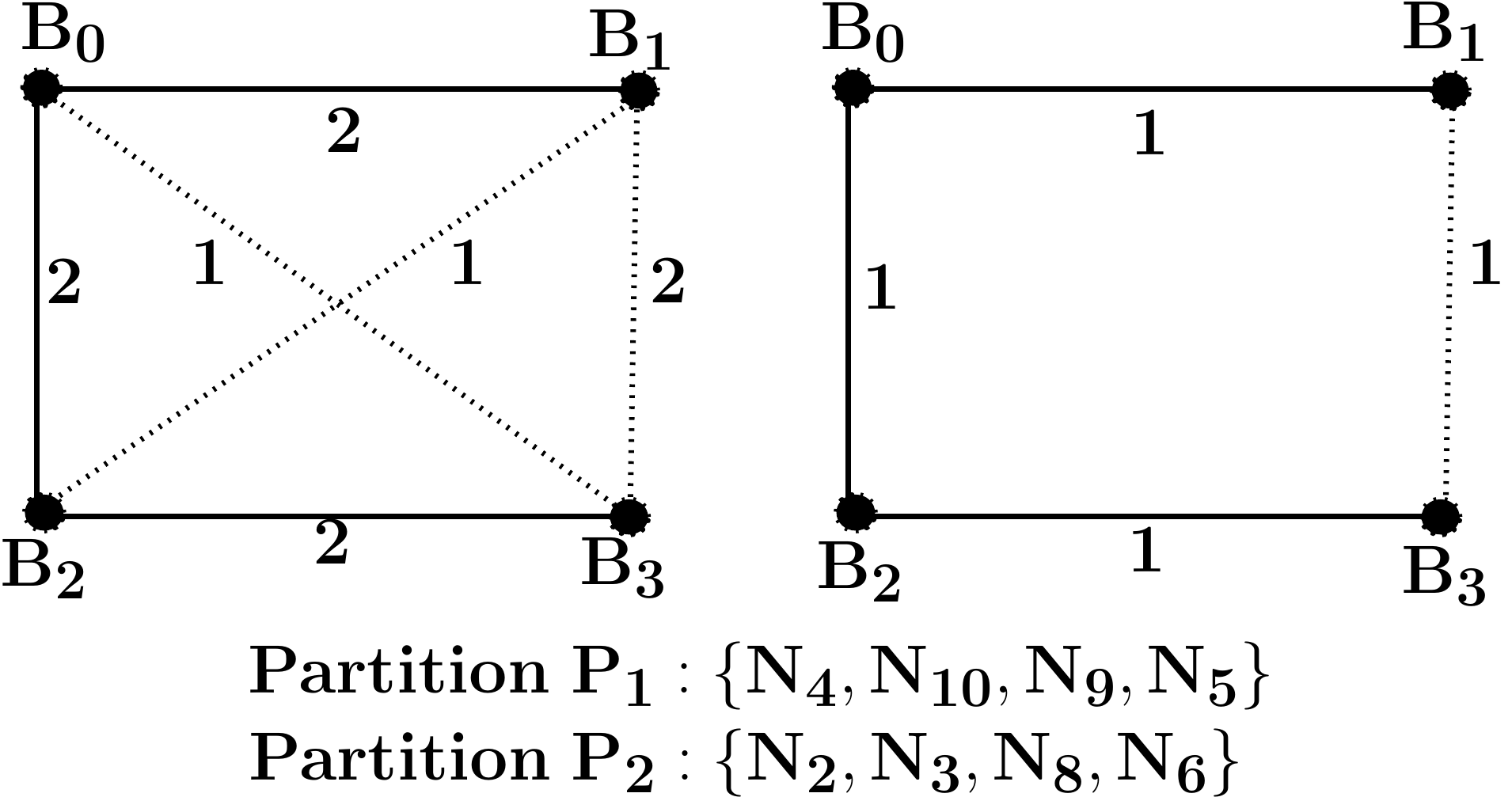}
	    \caption{\scriptsize{Two MST's from $G_{\cal B}({\cal B},E_{\cal B})$ of Figure \ref{fig4}}}
	     \label{partition_result}
	      \end{minipage}
	      \hspace{0.5cm}
	      \begin{minipage}[b]{0.4\linewidth}
		  \centering
		   \includegraphics[scale=0.22]{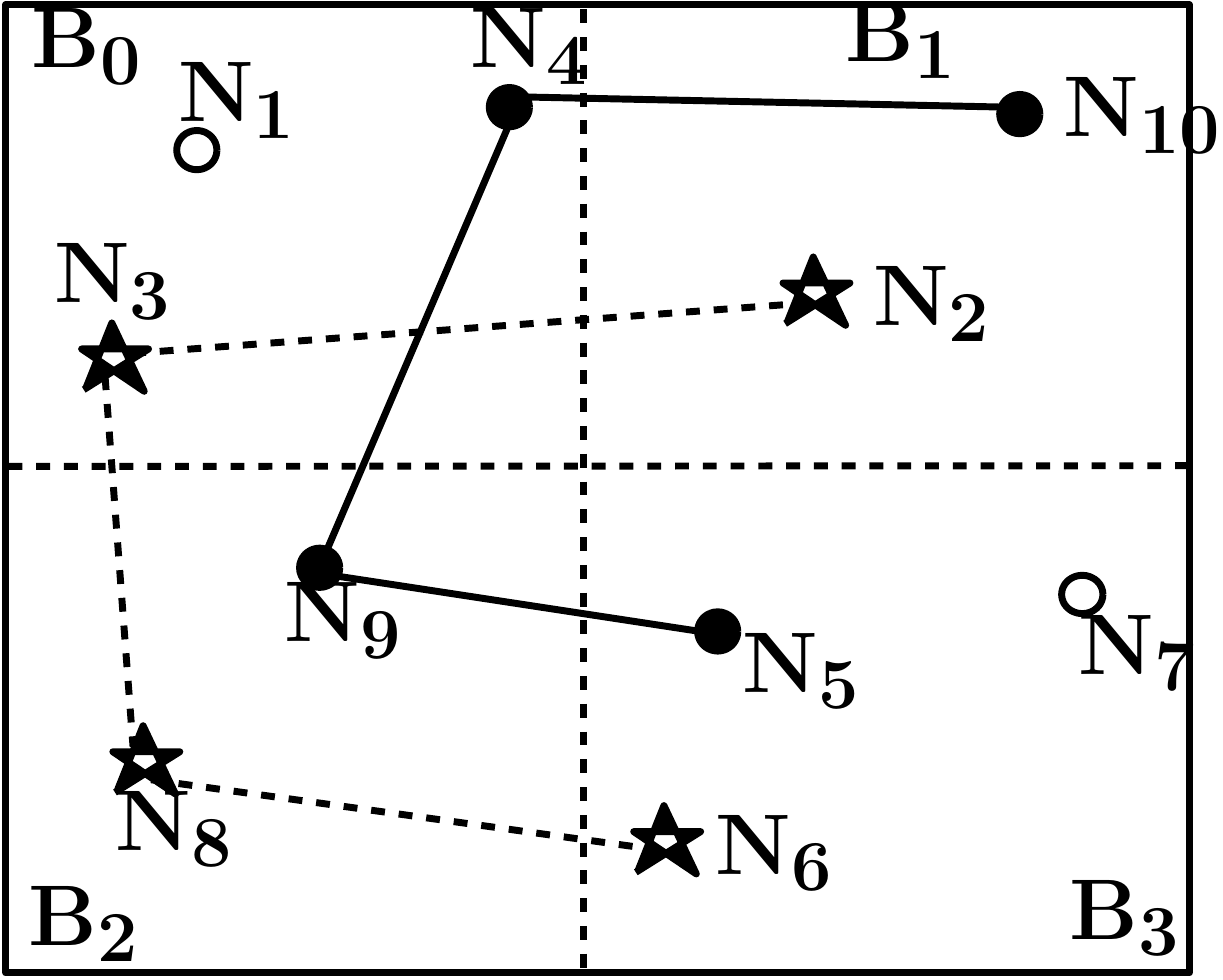}
	    \caption{\scriptsize{Connected node sets in two partitions covering $\cal Q$}}
	    \label{active_partition}
		\end{minipage}
		\end{figure}

{\bf {\it Step 1-} Construction of the Weighted Cellular Graph $G_{\cal B}({\cal B},E_{\cal B})$:} From ${{\it G}}({\cal N}, E)$, for each pair of adjacent blocks $B_i$ and $B_j$, a bipartite graph ${{\it G}}_D(B_i, B_j, E_D)$ is constructed where $E_D$ includes all bipartite edges existing between the pairs of nodes, one in $B_i$ and another in $B_j$ in ${{\it G}}({\cal N}, E)$. Next to find the edge weights of $G_{\cal B}({\cal B},E_{\cal B})$, i.e., the number of disjoint edges between blocks, for every adjacent block pairs we model the problem as the {\it maximum~flow} problem and apply the {\it Ford-Fulkerson} algorithm. From ${{\it G}}({\cal N},E)$ of Figure \ref{fig2}, the maximum flow graph for blocks $B_0$ and $B_1$ is shown in Figure \ref{fordfulkerson}. A virtual source and a virtual sink node are added to the bipartite graph for each pair of blocks and each edge is assumed to have unit flow capacity. Hence, for each pair of blocks the maximum flow gives the number of disjoint edges between the 
blocks.Figure \ref{fig3} shows the list of disjoint edges existing between all pairs of blocks of ${{\it G}}({\cal N}, E)$ obtained by this method. Figure \ref{fig4} shows the corresponding weighted cellular graph $G_{\cal B}({\cal B},E_{\cal B})$.
 \\{\bf {\it Step 2-} Formation of the partition list:} The graph $G_{\cal B}({\cal B},E_{\cal B})$ is stored as a $({\cal B}\times{\cal B})$ adjacency matrix, say {\it BlockMatrix} where {\it BlockMatrix(i, j)} is the weight of the edge (${\cal B}_i, {\cal B}_j$) in $G_{\cal B}({\cal B},E_{\cal B})$. Next the maximum spanning tree ($\cal MST$) is constructed following the $Prim's$ algorithm replacing each edge weight $W_{ij}$ as $-W_{ij}$. Each edge of the $\cal MST$ connects a pair of sensor nodes belonging to different blocks. Once we get an $\cal MST$, all selected pairs of sensor nodes are included in the partition and all the corresponding incident edges are deleted from ${{\it G}}({\cal N}, E)$. The $disjoint\ edge$ list and the edge weights are updated in $G_{\cal B}({\cal B},E_{\cal B})$. This procedure is repeated until it fails to output an $\cal MST$. The nodes contained in each spanning tree is a desired connected set cover.\\
 Execution of this algorithm on ${{\it G}}({\cal N}, E)$ of Figure~\ref{fig2} results two spanning trees shown in Figure~\ref{partition_result}. Figure~\ref{active_partition} shows the corresponding set covers partition.
		
	\vspace{-0.2cm}
{\bf  Centralized algorithm: CCSP }
   \begin{algorithm}[ht]
         \scriptsize
        \SetLine
\caption{\scriptsize{Centralized algorithm: CCSP}}
      \KwIn{ ${{\it G}}({\cal N}, E)$, $\cal B$, block -ID of each node $N_i\in N$}
      \KwOut{Partition list $\{P_i\}$}
       \For {each pair of blocks $B_i, B_j \in {B}$}{
	   Apply {\it Ford-Fulkerson} algorithm to calculate the edge weight $W_{i, j}$ and insert in the {\it BlockMatrix} \;
       Find the node pairs corresponding to each edge and insert in the {\it disjoint edge list};// construction of $G_{\cal B}({\cal B},E_{\cal B})$.       
	}
	$i=0$\;
       \While{ the $\cal MST$ by Prim's algorithm exists in $G_{\cal B}({\cal B},E_{\cal B})$ }
        {
	  $i=i+1$\;
               Insert in $P_i$ the node pairs for each edge ($i, j$) in $\cal MST$\;
	       Delete it from the {\it disjoint edge list} and modify $W_{i, j}$ and the {\it BlockMatrix} \;
	      }
  \normalsize
  \end{algorithm}
  \vspace{-0.6cm}
 \paragraph{\bf Correctness and Complexity Analysis} 
 By construction, the edge weight $W_{ij}$ of an edge $(B_i, B_j)$ in the Weighted Cellular Graph $G_{\cal B}({\cal B},E_{\cal B})$ obtained by $Ford-Fulkerson$ method essentially gives the number of disjoint edges existing between block pairs $B_i$ and $B_j$. Now it is evident that any spanning tree of the graph $G_{\cal B}({\cal B},E_{\cal B})$ contains at least one sensor node from each block, i.e. it covers ${\cal Q}$. Since all the sensor nodes within the same block are connected, the node set selected by the edges of the spanning tree is also connected. In {\it step 2} of the algorithm {\it Maximum Spanning Tree} was constructed in a greedy manner in an attempt to extract maximum number of spanning trees from $G_{\cal B}({\cal B},E_{\cal B})$. It proves the correctness of the algorithm. \\ 
 Let $n$ sensor nodes be deployed over $m$ number of blocks following a uniform random distribution. Therefore, the maximum number of edges possible between each pair of adjacent blocks is {\it O}$(n^2)$, and the maximum possible flow can be {\it O}$(n)$. Hence the time complexity in computing the edge weight for a pair of blocks by {\it Ford-Fulkerson} method becomes $O(n^3)$. For all pairs of adjacent blocks it needs $O(n^{3}m^{2})$ time in {\it Step 1}. In {\it Step 2}, to construct the adjacency matrix of graph ${{\it G}}_{\cal B}({\cal B},E_{\cal B})$ it requires $O(m^2)$ time. From this graph, generation of each partition by $Prim's$ algorithm takes $O(m^2)$ time. Since the maximum number of partitions possible is $\frac{ n}{m}$, in the worst case, total time needed in {\it Step 2} is $O(m^2+n.m)$.
 Therefore, the time complexity of the centralized algorithm is $O( n^3+ m^2+ n.m)$, i.e. $O( n^3)$, for $m<<n$.
\vspace{-0.3cm}
\subsection{\bf Distributed Algorithm: DCSP}
For a large self-organized network, it is difficult for a sensor node to gather the information about the whole network since it demands excessive overhead in terms of message communication. Also, the centralized algorithms are computation intensive and not feasible. In applications with wireless sensor networks, distributed algorithms are suitable where nodes may take decisions based on their local information only. The centralized algorithms though infeasible can act as the benchmarks for the evaluation of the performance.\\
Let $n$ number of sensor nodes be randomly distributed over the query region ${\cal Q}$ which is divided into $m$ number of blocks, each node and each block have their unique ID. In the distributed algorithm, we assume that depending on node density, a certain fraction of nodes select themselves as the {\it leader} nodes randomly. The predetermined value of the global variable {\it Leader Probability} $(L_P)$ is broadcasted initially to all nodes.\\
The following types of messages are communicated among the nodes during the execution of the algorithm. 
\vspace{-0.5cm}
\begin{itemize}
\item $SelectReq(L_i, i, k)\ message$: Node $i$ selects its neighbor $k$ for inclusion in its partition and sends a request message to Leader $L_i$ via multihop path.
\item $include(L_i, k)\ message$: The leader $L_i$ initiates the message to include node $k$ in its partition which is broadcasted to all members of the partition.
\item $confirm(L_i, k)\ message$: The node $k$ confirms that it is joining the partition of leader $L_i$.
\end{itemize}
\vspace{-0.5cm}
 Here follows an outline of the distributed algorithm.
 \begin{itemize}
 \vspace{-0.4cm}
 \item {\bf Phase 1 - Initialization :}
 
           Each node $N_i $ initializes {\it node-Id}, {\it block-Id}, {\it Leader~probability~$L_p$}, {\it block-status $B(N_i)$}(covered or not), $flag(N_i)$ (already joined a partition or not), $degree \ {\cal D}$ and neighbor list $NL(i)$.
\item{\bf Phase 2 - Leader election :} 
    
      Each node $N_i\in {\cal N}$ generates a random number $R$ where $0\leq R\leq 1$. If $R \leq L_p$, the node $N_i$ becomes a $Leader$ node. Each leader node $L_i$ sets $flag(L_i)=1$, $block$-$status\ B(L_i)=1$ and $parent(L_i)=\emptyset$.
\item{\bf Phase 3 - Selection and Confirmation }

	 Each sensor node $N_i$ in a partition $P_i$ selects a node $k$ from its $NL(i)$ and the nodes selected by its children such that it belongs to a block yet to be covered and has minimum degree, if it exists, else selects the neighbor with minimum degree ${\cal D}$, else $k= invalid~ID$ and forwards {\it SelectReq message}.\\   
	If  $parent=\emptyset$, the node sends an $include(L_i, k)$ message to the selected node $k$.
        
A node $k$ receiving $include(L_i,k)$ messages sends {\it confirm ($L_i$,k)} message to the sender node which have minimum degree ${\cal D}$ and updates its $flag$ and $block$-$status$ an
$parent$. Each node receiving a $confirm$ message forwards it to its parent until parent=$\emptyset$. If $parent=\emptyset$, include $k$ in $P_i$ and repeat until ${\cal Q}$ is covered or all nodes in a partition fail to select a node.
\end{itemize}
 { \bf Distributed algorithm: DCSP}
 \begin{algorithm}[!h]
  \scriptsize
  \SetLine
\caption{\scriptsize{ Distributed algorithm: DCSP}}
 \KwIn{1-hop neighbor list of each node $NL(i)$ with degree ${\cal D}$, {\it Block-Id}, {\it Block~status}, $flag$, $Leader Probability: L_P$}
 \KwOut{Partition $P(L_i)$ from leader $L_i$ }
  \For {each node-$i$}
{
    {\bf Step 1.} Initialize $flag(i) \leftarrow 0$, $L \leftarrow 0$, $Block\_status(i)\leftarrow 0$, $parent(i) \leftarrow \emptyset$\;
    {\bf Step 2.} Generate a random number $R$, if $R<L_P$ then $flag(i)\leftarrow 1$, block status$\leftarrow 1$ and $L\leftarrow 1$, $P(L_i)\leftarrow L_i$\; 
    {\bf Step 3.} if $L=0\ and\ flag(i) = 0$ wait and listen to messages from 1-hop neighbors\;
    {\bf Step 4.} if $flag(i)=1$ and received all $SelectReq$ messages from its children in $P(L_i)$ then selects node-$j$ that covers a new block and/or has the minimum ${\cal D}$ from $NL(i)~\cup$ nodes in $SelectReq$ messages from its children; if none is found $j=0$\;
    {\bf Step 5.} node-$i$ forwards $SelectReq(L_i, N_i, j)$ to $parent(i)\neq \emptyset$ else if $j=0$ broadcast 'failed' message and terminate else send $include$ message to node-$j$ \;
    {\bf Step 6.} if node-$i$ receives $include$ messages and $flag(i) = 0$ then sends a $confirm$ message to the parent node-$k$ with minimum degree ${\cal D}$\; update $flag(i)\leftarrow 1$, $Block$-$status(i)\leftarrow 1, parent(i)\leftarrow k$, $P(L_i)$, $NL(i)$\;
    {\bf Step 7.} if node-$i$ with $flag(i)=1$ receives a $confirm$ message then broadcasts it and updates $P(L_i)$; If $L=1$ and all blocks are covered broadcasts 'successful' message and terminates \;
 }
 \normalsize
 \end{algorithm}
\vspace{-0.5cm}
\paragraph{\bf Correctness and Message Complexity} From the outline of the distributed algorithm, it is evident that in phase 3, starting from each leader, each node of a partition attempts to include one of its neighbors and thus grows the partition. Hence by construction each partition remains connected. Also, in each iteration of phase 3, a partition includes a new node to cover a block so far uncovered whenever possible, and the process terminates either when all blocks are covered and it results a successful partition, or when all nodes in a partition fails to select nodes. Hence, it is clear that the algorithm results connected partitions which cover ${\cal Q}$.\\
In the distributed algorithm, let in the $k^{th}$ iteration, $(k+1)^{th}$ node is to be selected for a partition $P_i$.
For the new node discovery ($(k-1)$) $SelectReq$ messages are transmitted, one from each node in $P_i$, except the leader. Also, $k$ $include$ messages and $k$ $confirm$ messages are to be transmitted to include the new node. Hence in $k^{th}$ iteration, total number of message transmission is $(3k-1)$.
In the worst case, all the nodes may get included in one partition and the message complexity becomes $\displaystyle\sum\limits_{k=1}^{n-1} (3k-1)$, i.e. $ O(n^2)$. Hence per node average message complexity is $O(n)$. However, in each iteration, each node transmits only $3$ messages, and in the best case it may require $m$ rounds to cover $m$ blocks, whereas in the worst case it may require $n$ rounds, where $m << n$. Still compared to the dynamic protocols where each node transmits in some intervals during the whole lifetime of the network, it is expected that the proposed technique will be able to manage with much less message overhead since the computation is to be done just once. The fact is also validated by simulation results.
\subsection{Fault Recovery}
Since the proposed distributed algorithm is executed just once during the initialization of the network, it is challenging to reconstruct the partitions in case of failures. In this paper, we focus on node faults due to complete energy depletion. If any node in an active partition relinquishes its energy completely the partition may fail to cover the query region and also it may get disconnected. To recover from such failures, a distributed algorithm is proposed. Here, each active node-$i$ checks if its energy level goes below a threshold level when it broadcasts a {\it failed} message with the node-ID's of its parent node $parent(i)$ and its children nodes in its partition. In response to it the node $parent(i)$ acts as a leader node and follows the procedure {\it DCSP} to cover the block of node-$i$ and the blocks of its children nodes by including some additional nodes from the set of free nodes which are not included in any partition.
Since the procedure ensures that included nodes are connected with $parent(i)$ and the nodes in the same block are also connected, the additional nodes can successfully cover all the blocks, making the partition connected. In the worst case, if it is not successful, the partition is dismantled and all its nodes are declared free. If the maximum degree of a node be $D$, it may require $D$ iterations at most causing {\it O$(D^2)$} messages to repair the partition in case of a single fault.
\section{Simulation Results and Discussion}
\label{sec_4}
\begin{figure}[ht]
		\begin{minipage}[b]{0.49\linewidth}
		  \centering
		  \includegraphics[scale=0.37]{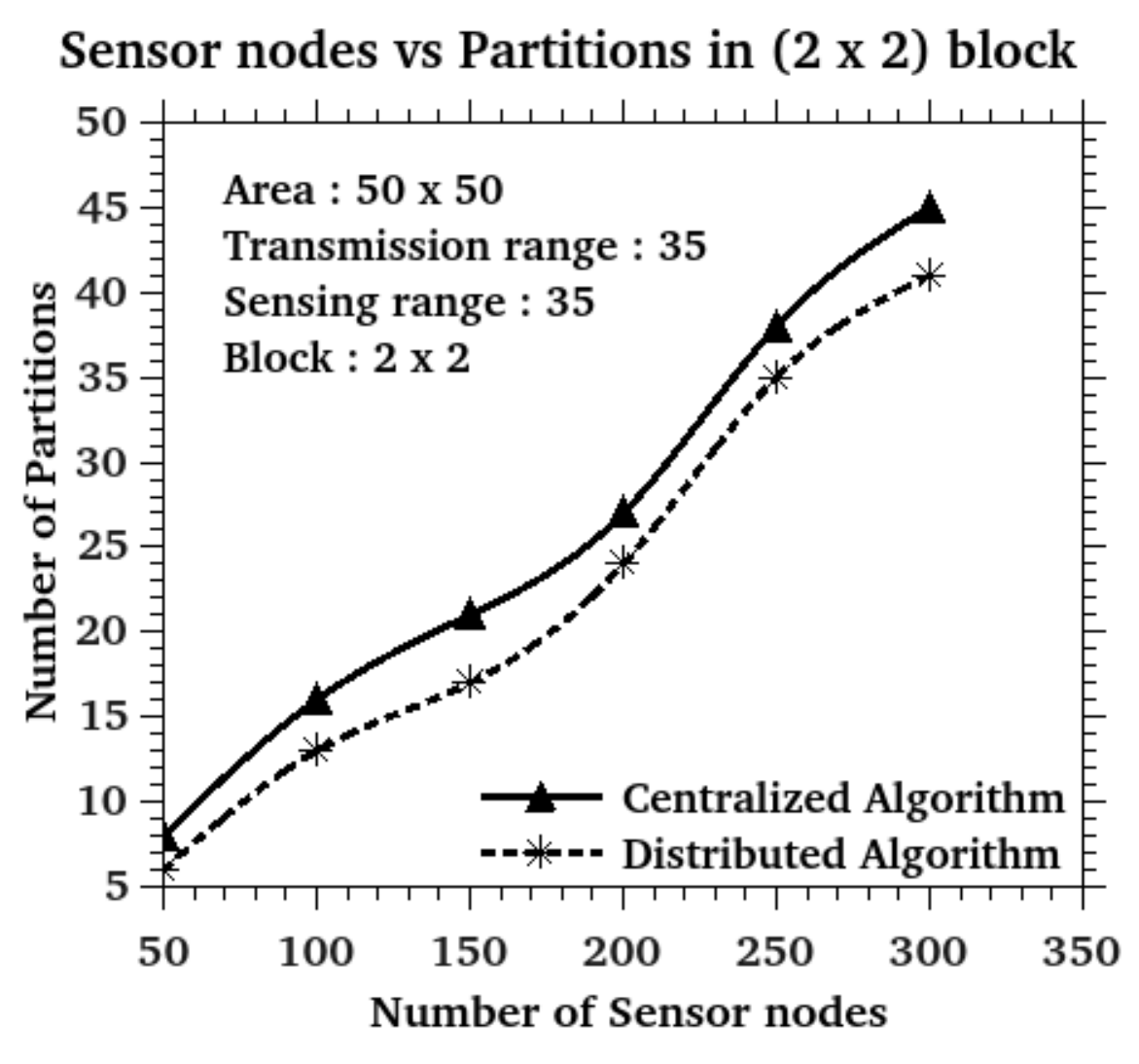}
		  \caption{\scriptsize{Nodes vs partitions for $2\times2$ of blocks}}
		   \label{2x2}
	      \end{minipage}
 	      \hspace{0.1cm}
	      \begin{minipage}[b]{0.49\linewidth}
		  \centering
		  \includegraphics[scale=0.31]{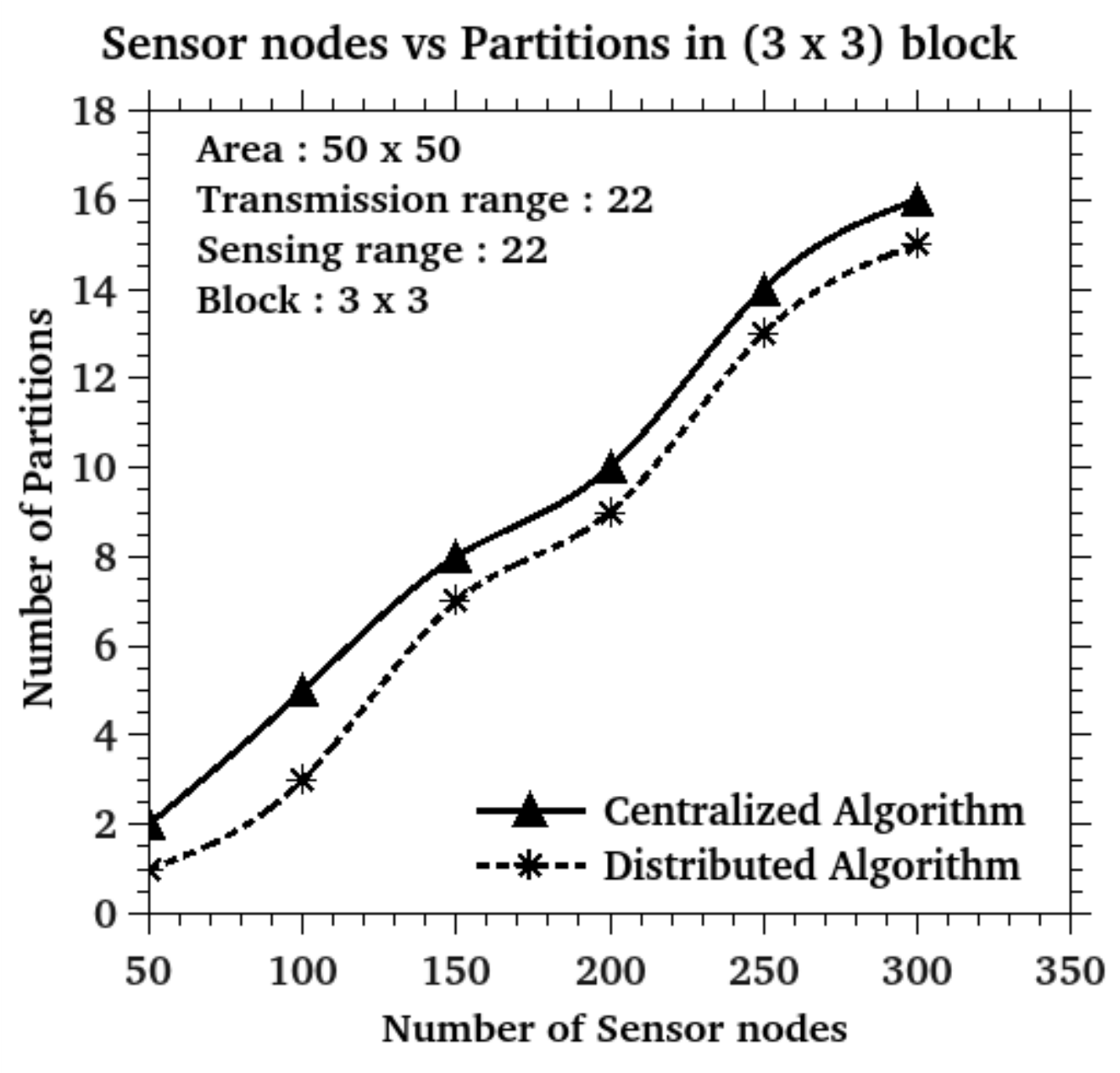}
		  \caption{\scriptsize{Nodes vs partitions for $3\times 3$ of blocks}}
		   \label{3x3}
	      \end{minipage}
\end{figure}

\begin{figure}[ht!]
		\begin{minipage}[b]{0.40\linewidth}
		  \centering
		  \includegraphics[scale=0.35]{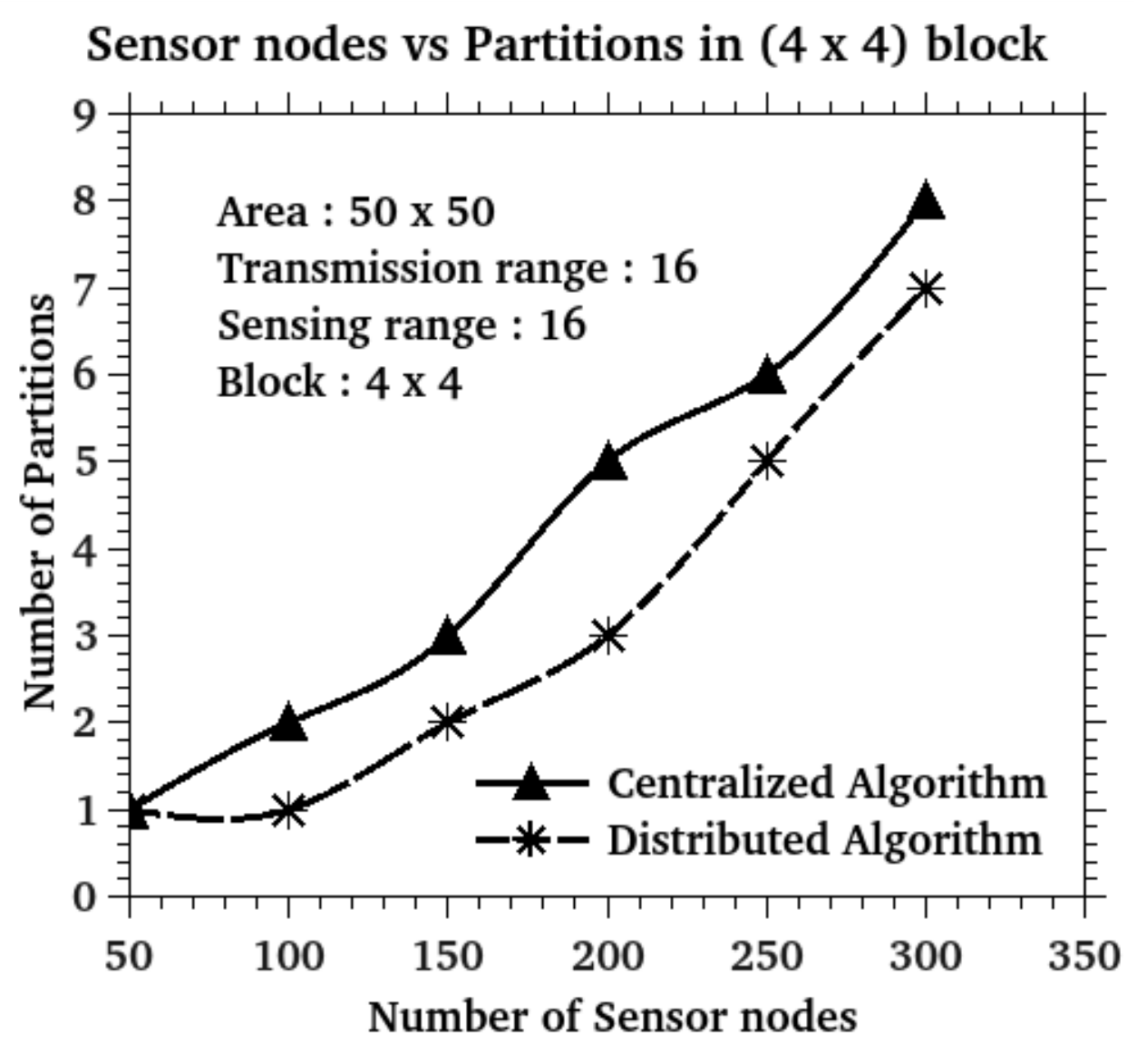}
		  \caption{\scriptsize{Nodes vs partitions for $4\times 4$ of blocks}}
		   \label{4x4}
	      \end{minipage}
  	      \hspace{0.82cm}
	      \begin{minipage}[b]{0.40\linewidth}
		  \centering
		  \includegraphics[scale=0.32]{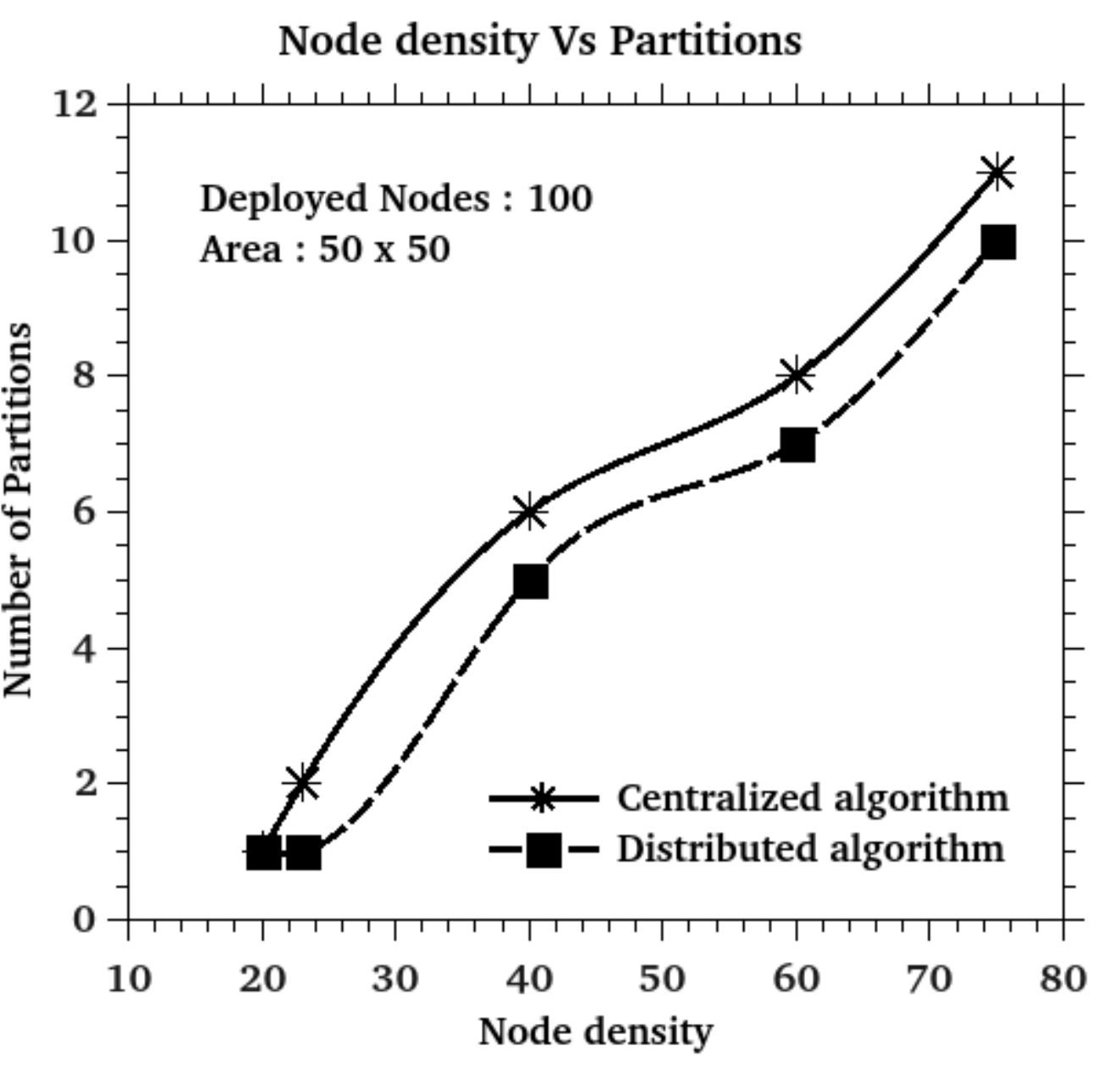}
	    		  \caption{\scriptsize{Node density vs number of partitions}}
		 \label{fig_densityvspartition}
		\end{minipage}
\end{figure}

To evaluate the performance of the proposed distributed algorithm, we have done simulation studies on network simulator NS-2.34. For simulation, $50 \leq n \leq 350$ sensor nodes are deployed randomly over a $50\times50$ unit query region.\\
The centralized and the distributed algorithms are executed on same network and the experiment is repeated $20$ times for each setting. The results are shown in Figures~\ref{2x2}-\ref{4x4} for ($2\times 2$), ($3\times 3$) and ($4\times 4$) grid. 
It has been found that though the distributed version uses much less computation and communication, the number of successful partitions generated by it is almost comparable with that achieved by the centralized algorithm, especially for larger block sizes. 
	     \begin{figure}[ht!]
		  \begin{minipage}[b]{0.425\linewidth}
		  \centering
		   \includegraphics[scale=0.30]{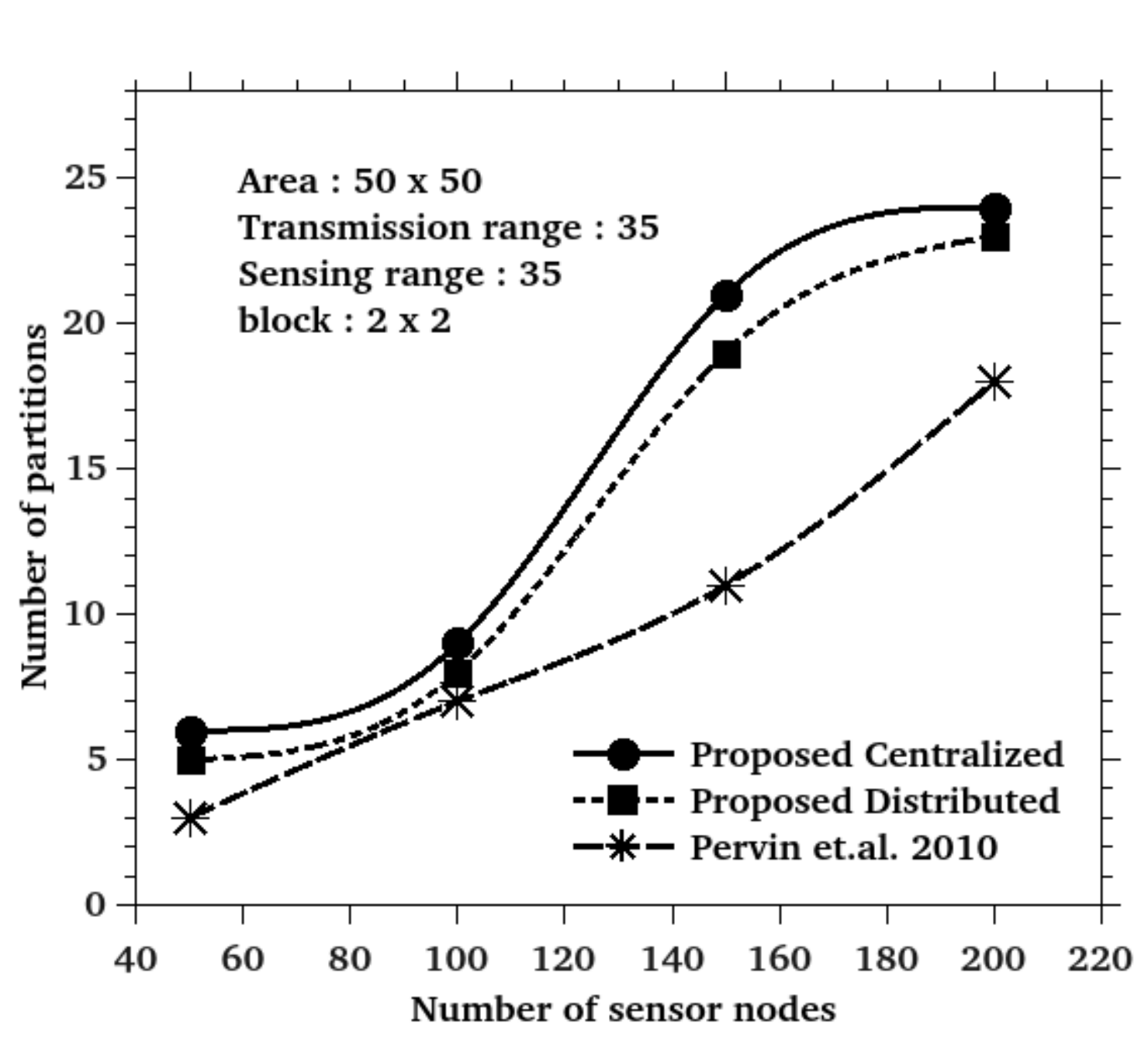}
		  \caption{\scriptsize{Comparison with \cite{Pervin} in terms of number of partitions }}
		  \label{fig7}
	      \end{minipage}
	      \hspace{0.8cm}
	      \begin{minipage}[b]{0.45\linewidth}
		  \centering
		  \includegraphics[scale=0.256]{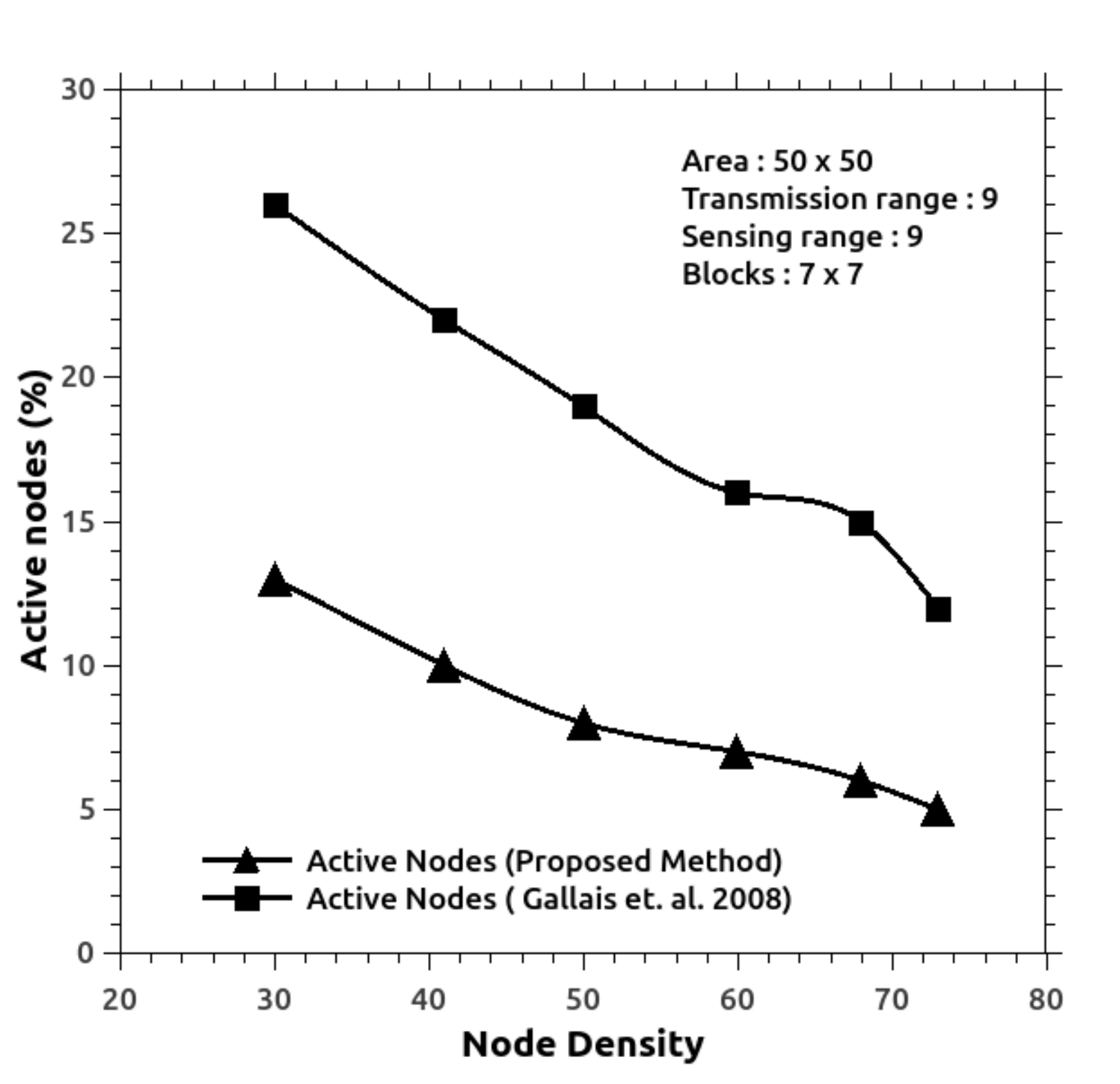}
		  \caption{\scriptsize{Comparison with \cite{Gallais} in terms of to the active nodes in a round}}
		  \label{compare_with_gallais_active_nodes}
	      \end{minipage}
	     \begin{minipage}[b]{0.40\linewidth}
		  \centering
		    \includegraphics[scale=0.30]{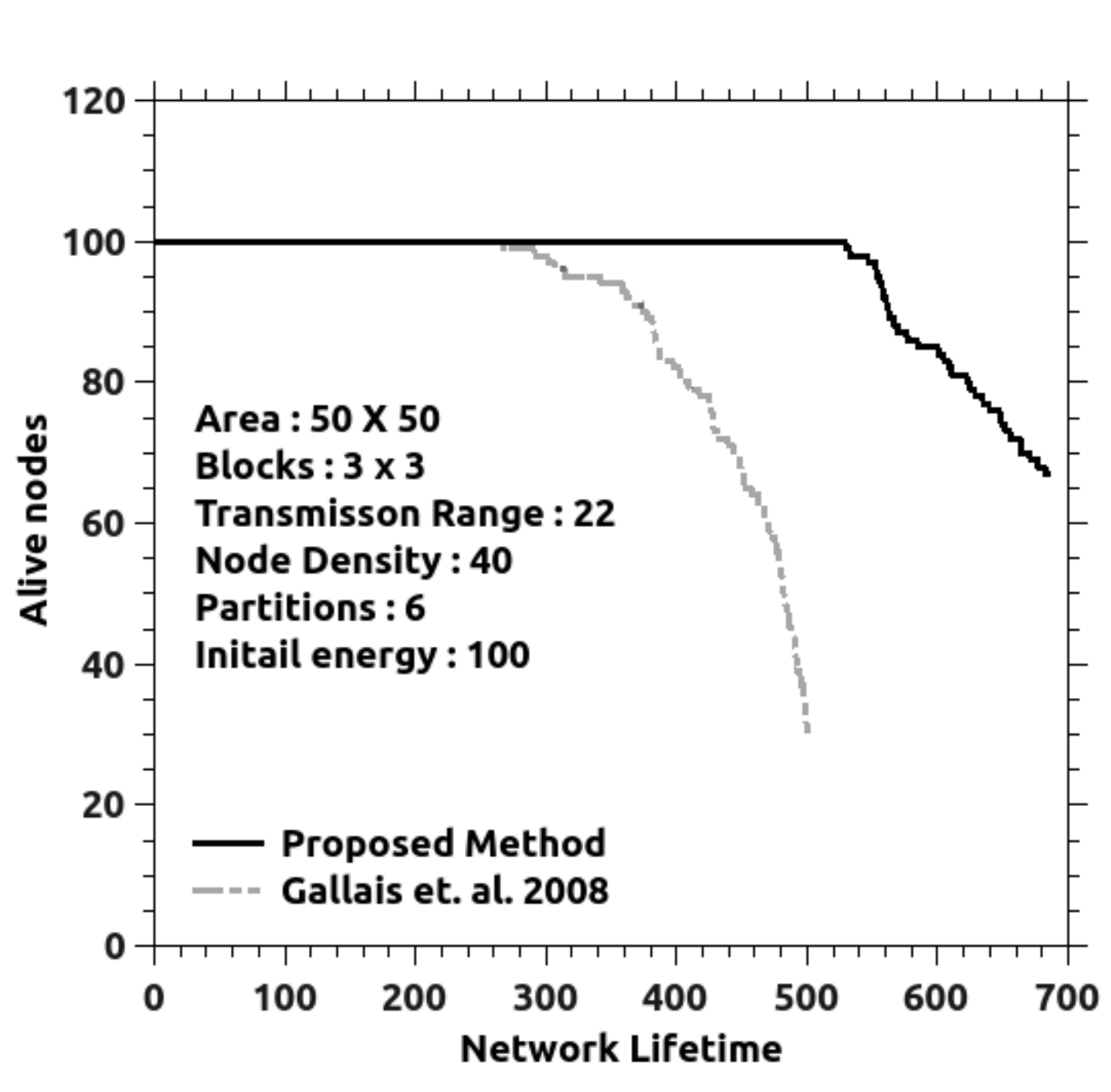} 
		  \caption{\scriptsize{Comparison with \cite{Gallais} in terms of network lifetime }}
		  \label{compare_with_gallais_lifetime}
		\end{minipage}
		      \hspace{0.8cm}
		\begin{minipage}[b]{0.46\linewidth}
		  \centering
		  \includegraphics[scale=0.30]{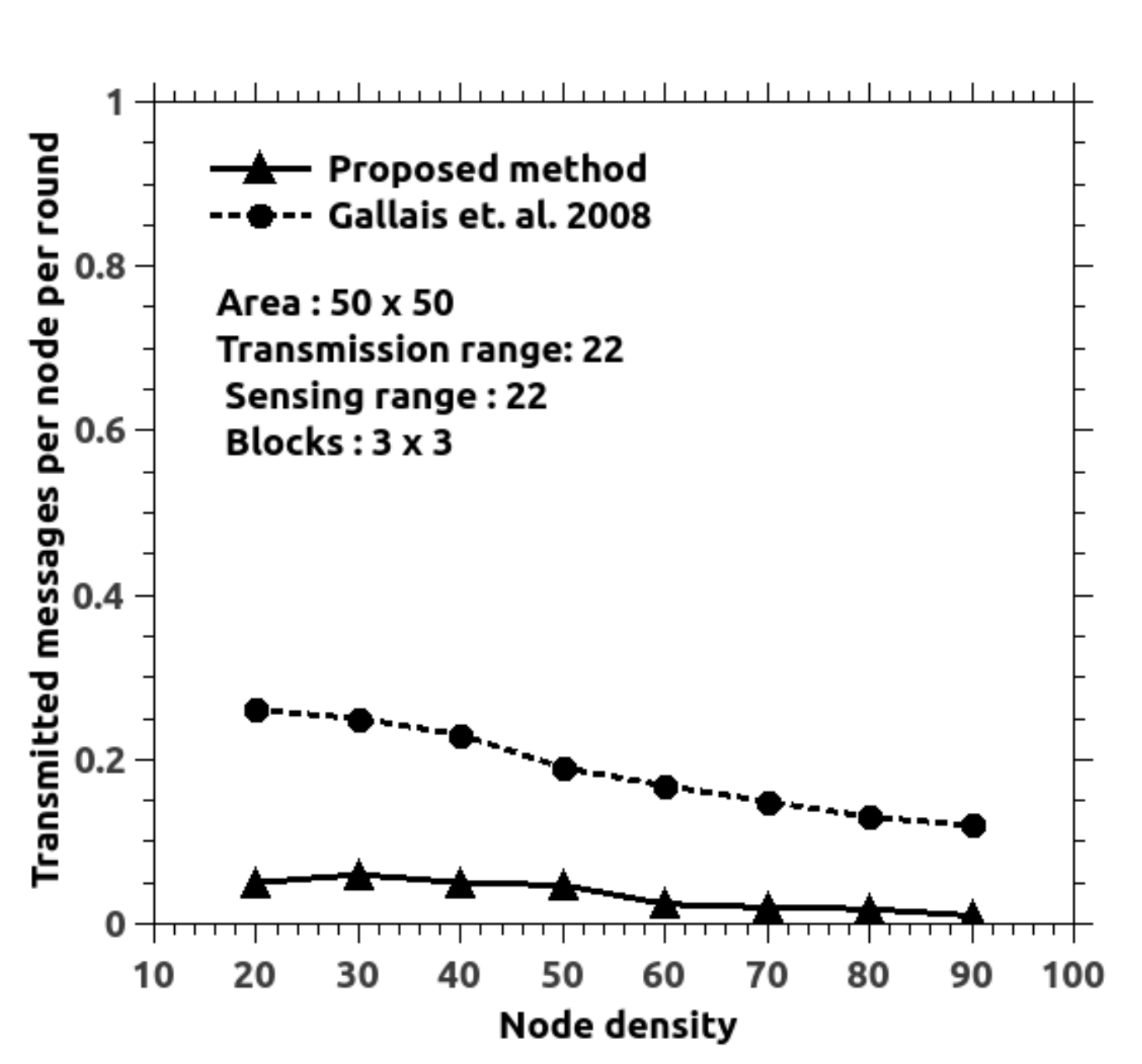}
		  \caption{\scriptsize{Comparison with \cite{Gallais} on transmitted messages per node per round}}
		  \label{compare_with_gallais_messages}
	      \end{minipage}
		\end{figure}
 Figure~\ref{fig_densityvspartition} shows how the number of partitions increases with network density, where the average degree ${\cal D}$ of all the active sensor nodes is considered to be the network density.
 We also have compared the performance of the proposed algorithms with the distributed algorithm proposed in \cite{Pervin}. Under $100\%$ coverage criteria, varying $n$ within the range $50 \leq n \leq 300$, the number of partitions are shown in Figure~\ref{fig7} for $T=35 units$. It reveals the fact that under the same conditions, the proposed distributed algorithm performs better in terms of number of partitions. Also, the improvement becomes significant in case of over deployed networks. For example, with $150$ nodes the number of partitions by our algorithm is almost double than that resulted in \cite{Pervin}.
Figure~\ref{compare_with_gallais_active_nodes} shows that the proposed algorithm in terms of percentage of active nodes performs significantly better than the dynamic scheduling protocol presented in \cite{Gallais} ({\it AO technique}). In Figure~\ref{compare_with_gallais_lifetime} network lifetime is compared under the fault model discussed in Section 4. It shows that {\it DCSP} algorithm improves the lifetime significantly. Finally  Figure~\ref{compare_with_gallais_messages} shows in terms of number of transmitted messages per node per round, the {\it DCSP} performs much better compared to \cite{Gallais}.
\\In summary, the performance comparison study establishes that our proposed algorithm outperforms \cite{Pervin} with respect to number of partitions and computation complexity and compared to \cite{Gallais}, it performs better in terms of percentage of active nodes, message complexity and lifetime.

\section{Conclusion}
\label{sec_5}
In this paper, we have presented a feasible solution to enhance the lifetime of a wireless sensor network. To maximize the lifetime of the WSN, we have addressed the {\it Connected Set Cover Partitioning problem} for finding maximum number of mutually exclusive connected sets of sensor nodes with required coverage for a given query region. If we obtain $\cal P$ such covers and one set is activated every $\cal P$ time interval, the life time of the network can be enhanced $\cal P$ times. We propose an $O(n^3)$ time $centralized$ algorithm and its $distributed$ version both to be executed just once during the initialization. Simulation studies show that the performance of the distributed algorithm is comparable with the centralized one though the former one requires much less computation and less message overhead. Comparison with existing distributed protocols \cite{Pervin} and \cite{Gallais} shows that the proposed algorithm without the knowledge of exact locations of the nodes performs better either in terms 
of number of partitions, or in terms of active nodes, communication overhead and network lifetime.

 \bibliographystyle{acm}
\bibliography{reff}
\end{document}